\documentclass[aps,prl,reprint,groupedaddress,amsmath,amssymb]{revtex4-2}
\usepackage{graphicx}
\usepackage{xcolor}
\usepackage{hyperref}

\definecolor{ao(english)}{rgb}{0.0, 0.5, 0.0}

\graphicspath{{./}{Figures/}}

\begin{document}

\title{Global Kinetic Simulations of Monster Shocks and Their Emission}

\author{Dominic Bernardi}
\email{bdominic@wustl.edu}

\author{Yajie Yuan}
\email{yajiey@wustl.edu}
\author{Alexander Y. Chen}
\email{cyuran@wustl.edu}
\affiliation{Physics Department and McDonnell Center for the Space Sciences, Washington University in St.\ Louis; MO, 63130, USA}

\date{\today}

\begin{abstract}
Fast magnetosonic waves are one of the two low-frequency plasma modes that can exist in a neutron star magnetosphere. It was recently realized that these waves may become nonlinear within the magnetosphere and steepen into some of the strongest shocks in the universe. These shocks, when in the appropriate parameter regime, may emit GHz radiation in the form of precursor waves. We present the first global Particle-in-Cell simulations of the nonlinear steepening of fast magnetosonic waves in a dipolar magnetosphere, and quantitatively demonstrate the strong plasma acceleration in the upstream of these shocks. In these simulations, we observe the production of precursor waves in a finite angular range. Using analytic scaling relations, we predict the expected frequency, power, and duration of this emission. Within a reasonable range of progenitor wave parameters, these precursor waves can reproduce many aspects of FRB observations.

\end{abstract}

\maketitle

\paragraph{Introduction.---}Magnetars are a class of young neutron stars with super-strong magnetic field, which can reach $10^{15}\,{\rm G}$ on the stellar surface~\cite{2017ARA&A..55..261K, 2015SSRv..191..315M}(we use cgs units throughout this Letter). The evolution of the strong internal field is believed to induce starquakes or crustal displacements that twist up the external field, injecting energy into the magnetosphere, which eventually powers the prolific X-ray activities we observe from these objects~\cite{1996ApJ...473..322T}. Magnetars may also be promising progenitors of fast radio bursts (FRBs)---mysterious, millisecond duration radio bursts of cosmological origin. In fact, a simultaneous X-ray and radio burst was observed from the galactic magnetar SGR~1935+2154~\cite{2020Natur.587...54C, 2020Natur.587...59B}, corroborating this connection. However, a detailed physical mechanism for such simultaneous emission is still under debate.

Extensive analytic and numerical work has demonstrated that dynamic activity around magnetars will inevitably produce energetic fast magnetosonic waves. These waves can be launched either directly at the stellar surface through starquakes~\cite{1989ApJ...343..839B, 2020ApJ...897..173B, Beloborodov2023, 2025arXiv250812567Q, 2025arXiv250818033B}, or further out in the magnetosphere through mode conversion~\cite{2021ApJ...908..176Y, 2024ApJ...972..139M, 2025ApJ...980..222B, 2025ApJ...987...42C}. They can also be generated during cataclysmic stellar collapse~\cite{2024ApJ...974L..12M}, or a merger event~\cite{2025ApJ...982L..54K}.

A fast wave emitted from near the neutron star expands spherically, and its amplitude decreases with radius as $\delta B\propto 1/r$. However, the background dipole magnetic field follows $B_\mathrm{bg}\propto 1/r^{3}$, therefore, $\delta B/B_{\rm bg}\propto r^2$---the relative amplitude grows with radius. If the wave amplitude is large enough, $\delta B/B_\mathrm{bg}$ may become order unity within the magnetosphere. In the high magnetization limit, the polarization of fast waves is such that the wave magnetic field $\delta\mathbf{B}$ is in the plane containing the wave vector $\mathbf{k}$ and the background magnetic field $\mathbf{B}_{\rm bg}$, while the wave electric field $\delta \mathbf{E}$ is perpendicular to this plane. Such a polarization implies that when $\delta B/B_\mathrm{bg}$ approaches unity, the wave magnetic field may cancel with the background magnetic field, potentially leading to macroscopic $E > B$ regions. In the MHD limit, such regions are not permitted, and the wave deforms to prevent $E > B$, converting a significant fraction of the wave energy into plasma bulk motion and forming a shock at every wavelength~\cite{2003MNRAS.339..765L,2022arXiv221013506C,Beloborodov2023}. These shocks are named ``monster shocks'' by~\cite{Beloborodov2023} due to the enormous Lorentz factor the upstream flow can reach. Recently, one dimensional kinetic simulations by~\cite{2025PhRvL.134c5201V} showed that these monster shocks can emit coherent precursor waves through the synchrotron maser instability, producing GHz signals that may resemble observed FRBs.

So far, existing studies of this monster shock have focused on the special case of the shock propagating perpendicularly to the magnetic field, which is only applicable on the equatorial plane of the magnetosphere. However, the nature of the shock and the observational consequences depend on the global geometry and field scaling. This type of problem is well-suited for global multidimensional Particle-in-Cell (PIC) simulations, as demonstrated by numerous works in the past decade, e.g., \cite{2014ApJ...795L..22C,2018ApJ...855...94P,2020A&A...642A.123C,2025ApJ...991...98S}. In this Letter, we present the first 2-dimensional global PIC simulations of the monster shock formation due to non-linear evolution of fast waves in a dipolar magnetosphere. For the first time, we demonstrate the scaling of the upstream Lorentz factor and its evolution in kinetic simulations with realistic geometry. We also show the self-consistent structure of the shock globally, and quantify how this structure affects the precursor wave production. Finally we estimate the properties of the resulting precursor waves and compare with the properties of observed FRBs.

\begin{figure*}[ht!]
    \centering
    \includegraphics[width=0.99\linewidth]{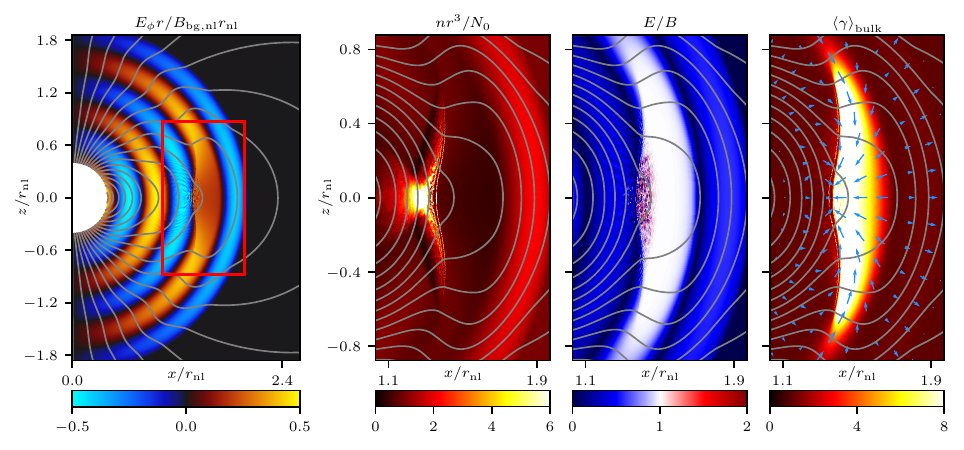}
    \caption{Global structure of the monster shock from our fiducial simulation with fast wave wavelength $\lambda=0.6r_{\rm nl}$ and $\sigma_{\rm bg, nl}=250$. The snapshot is taken at time $t=1.6r_{\rm nl}/c$. The left panel shows $E_{\phi}r/B_{\rm bg, nl}r_{\rm nl}$, where $B_{\rm bg, nl}$ is the equatorial magnetic field at the nonlinear radius. The subsequent three panels show a zoomed-in view of the region within the red box in the left panel, with colors representing the scaled plasma density $n r^3/N_0$ (where $N_0=n_{\rm bg}r^3$ is a constant), the ratio of the electric field to the magnetic field, and the Lorentz factor of the bulk flow, respectively. In all panels the gray lines are the magnetic field lines. In the rightmost panel, the blue arrows indicate the direction of the bulk flow, and the arrow lengths are proportional to the bulk velocity. Click \href{https://youtu.be/LyQCaSag0Zc}{here} for a YouTube video of these shocks.}
    \label{fig:Global}
\end{figure*}
\paragraph{Simulation Parameters.---}

Consider a magnetar with surface magnetic field $B_*=B_{15}10^{15}\,{\rm G}$ and stellar radius $r_*\sim 10^6\,{\rm cm}$, producing X-ray bursts with a typical luminosity $L=L_{42}10^{42}\,{\rm erg\,s}^{-1}$. Assuming an X-ray burst is produced from a fast wave with comparable luminosity, this wave will become nonlinear when $\delta B/B_{\rm bg}\sim 1/2$, at a radius
\begin{equation}\label{eq:r_nl}
    r_{\rm nl} = \sqrt{\frac{B_*r_*^3}{2}}\left(\frac{L}{c}\right)^{-1/4} \approx 2.9\times 10^8 \,B_{15}^{1/2} L_{42}^{-1/4} \,{\rm cm}.
\end{equation}
This is the central length scale of our problem. As a comparison, the light cylinder, where the magnetic field has to open up otherwise the plasma corotating with the neutron star would reach the speed of the light, is located at $r_{\rm LC}= cP/(2\pi)\sim4.8\times 10^9P_0\,{\rm cm}$, where $P=P_0 1\,{\rm s}$ is the spin period of the neutron star. In fact, as long as $L\gtrsim1.4\times 10^{37}P_0^{-4}B_{15}^2\,{\rm erg\,s}^{-1}$, the fast wave can become nonlinear within the light cylinder.

Our simulations are performed in 2D spherical coordinates $(r, \theta)$ assuming axisymmetry, using our PIC code \emph{Aperture}~\footnote{https://github.com/fizban007/Aperture4}. To capture the shock physics, the simulations need to resolve both the plasma skin depth $\lambda_p=c/\omega_p=\sqrt{m_ec^2/(4\pi n_{\rm bg} e^2)}$ and the gyration time scale $\omega_B^{-1}=m_ec/(eB_{\rm bg})$ at the nonlinear radius. The ratio of these two scales is the magnetization, which is a key parameter of this problem:
\begin{equation}
\sigma_{\rm bg,nl} = \frac{B_{\rm bg}^2}{4\pi n_{\rm bg}m_e c^2} = \left(\frac{\omega_B}{\omega_p}\right)^2.
\end{equation}
We have performed simulations with $\sigma_{\rm bg,nl}$ ranging from $160$ to $2000$, and fast wave wavelengths $\lambda$ from $0.2r_{\rm nl}$ to $r_{\rm nl}$. To capture the kinetic scale of the shock, we ensure at least 4 cells per skin depth at the non-linear radius in all our simulations. A detailed description of the simulation setup can be found in the Supplemental Material~\footnote{See Supplemental material for additional discussion which includes Refs. \cite{2008PhPl...15e6701V, 2001CoPhC.135..144E, 2025arXiv250304558C, 2015MNRAS.448..606C, chen-thesis, 2020ApJ...889...69C, 2017ApJ...844..133C, jackson1999classical, 2022ApJ...933..174Y, Zhang_2025}}.

\paragraph{Results.---}

Across a wide range of simulation parameters, we observe the formation of the monster shock near the nonlinear radius $r_{\rm nl}$. We are able to verify the condition of shock formation predicted by~\cite{2022arXiv221013506C} (see the Supplemental Material for quantitative discussion and demonstration). Figure~\ref{fig:Global} shows the global structure of the monster shock in our fiducial simulation with wavelength $\lambda = 0.6 r_{\rm nl}$, $\sigma_{\rm bg,nl}=250$, and $\lambda_p \approx 1.6\times 10^{-3}r_{\rm nl}$ at the nonlinear radius.  As the wave propagates beyond the nonlinear radius, the wave trough, in which $\delta B_\theta < 0$, significantly reduces the total magnetic field, because it partially cancels the background magnetic field. This first happens on the equator, then expands to higher latitudes as the fast wave propagates to larger radii. In this region (highlighted as a red box in the first panel of Figure~\ref{fig:Global}), the wave profile is significantly deformed to avoid $E > B$. As a result, a shock forms at the left edge of this region. In the upstream of this shock, the plasma bulk motion is accelerated in the $\mathbf{E}\times\mathbf{B}$ direction, reaching high Lorentz factors.

\begin{figure}
    \centering
    \includegraphics[width=0.99\linewidth]{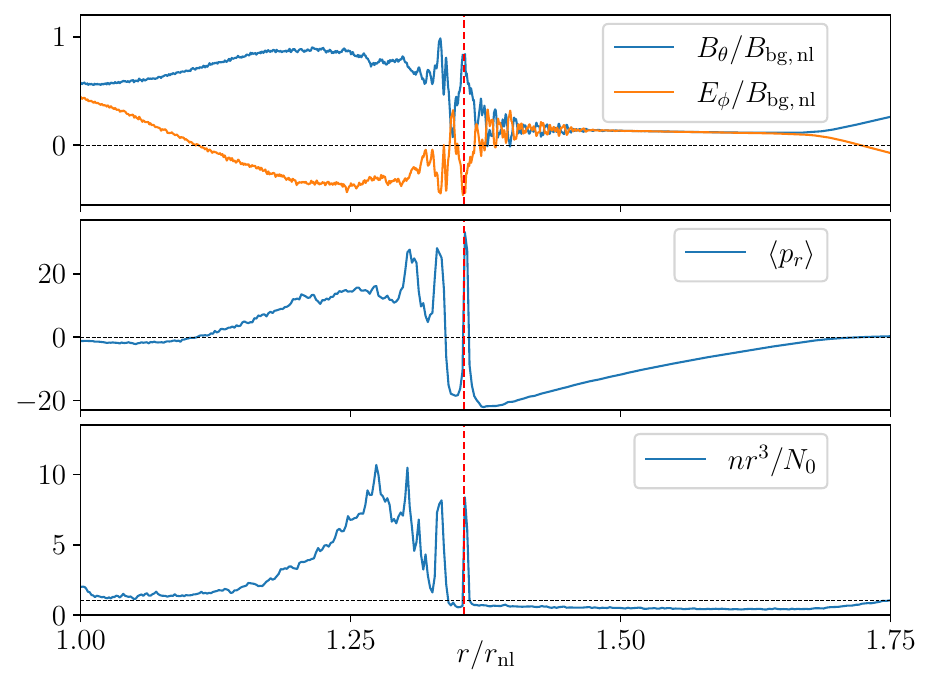}
    \caption{Structure of the shock on the equatorial plane in our fiducial simulation at the same time as in Figure~\ref{fig:Global}. The top panel shows $B_{\theta}/B_{\rm bg,nl}$ and $E_{\phi}/B_{\rm bg,nl}$. The middle panel shows the radial component of the average plasma momentum $\langle p_r\rangle$. The third panel shows the scaled plasma density $n r^3/N_0$ and the horizontal dashed line indicates the value of $N_0$. The red dashed line marks the location of the shock.}
    \label{fig:Equatorial}
\end{figure}

To see features of the shock more clearly, Figure~\ref{fig:Equatorial} shows a one-dimensional slice along the equator in the same fiducial simulation as in Figure~\ref{fig:Global}, taken at the same simulation time. The wave trough, as shown in $B_{\theta}$, is significantly deformed and flattened out. In this region, $E \approx B\approx B_{\rm bg}/2$, and the bulk flow accelerates almost linearly towards the $-\hat{r}$ direction, reaching a Lorentz factor of more than 20. Because the plasma is frozen into the magnetic field, the density also drops to half the background value~\cite{Beloborodov2023}. On the equatorial plane, the total magnetic field (in the $\hat{\theta}$ direction) is perpendicular to the bulk flow direction and the shock normal (both in the $\hat{r}$ direction); the resulting shock is a so-called perpendicular shock. In the shock transition region, two prominent solitons can be seen in all panels of Figure 2 at $r = 1.35r_{\rm nl}$ and $1.33r_{\rm nl}$ and as the striped feature extending to higher latitudes in the density panel of Figure~\ref{fig:Global}. The center of the leading soliton is indicated by the red dashed line in Figure~\ref{fig:Equatorial} and conventionally marks the shock front. Such solitons are characteristic of quasi-perpendicular collisionless shocks~\cite{1988PhFl...31..839A, 2025PhRvE.111d5209V}. The cavity formed by the two solitons produce precursor waves that are transmitted into the upstream~\cite{1992ApJ...391...73G, Plotnikov2019}, shown as the fluctuations with $E>B$ both in the top panel of Figure \ref{fig:Equatorial} and the third panel of Figure \ref{fig:Global}. To the left of the solitons, the plasma goes through further phase space mixing, heats up, and transitions into the downstream flow. We have checked that the shock jump condition is consistent with MHD theory in the no-cooling regime.

In an analytic calculation of monster shocks in the MHD framework, \cite{Beloborodov2023}~demonstrated that the plasma will be linearly accelerated to bulk Lorentz factors proportional to the background magnetization and the fast wave wavelength.
On the equatorial plane, the bulk Lorentz factor will evolve as a function of position:
\begin{equation}\label{eq:full_monster}
     \gamma_{u} \approx \frac{\sigma_{\rm bg, nl}}{2\pi} \frac{\lambda}{r_{\rm nl}}\left(\frac{r_{\rm nl}}{r}\right)^4\left(\pi - 2\arcsin\left(\frac{r_{\rm nl}}{r}\right)^2\right).
\end{equation}
$\gamma_{u}$ attains a maximum value of $\gamma_{\rm max} \approx 0.81\,\sigma_{\rm bg,nl}\lambda/(2\pi r_{\rm nl})$, at $r \approx 1.15r_{\rm nl}$.

\begin{figure}
    \centering
    \includegraphics[width=0.99\linewidth]{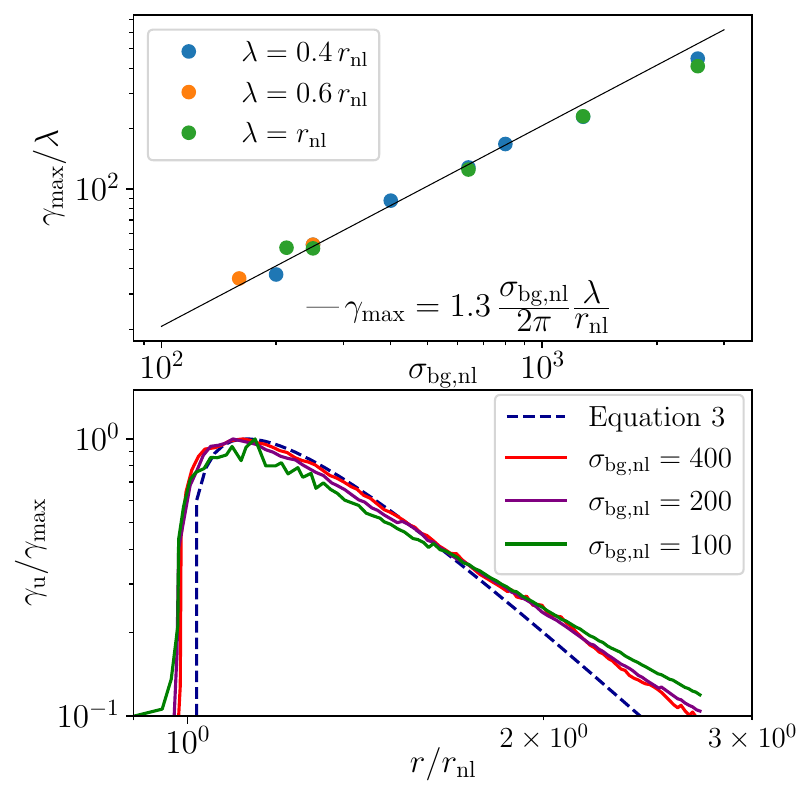}
    \caption{The top panel shows the scaling of the maximum upstream Lorentz factor $\gamma_{\rm max}$ on the equatorial plane with the background magnetization at the nonlinear radius $\sigma_{\rm bg,nl}$. We divide the measured $\gamma_{\rm max}$ by the fast wave wavelength $\lambda$. The resulting scaling relation is linear: $\gamma_{\rm max}/\lambda\propto \sigma_{\rm bg,nl}$. The bottom panel shows $\gamma_{u}$ as a function of position for simulations with different values of $\sigma_{\rm bg,nl}$, as well as the theoretically predicted curve, Equation~\eqref{eq:full_monster}. 
    For better comparison purposes, the curves are normalized by their own maximum value $\gamma_{\rm max}$.
    }
    \label{fig:Lorentz}
\end{figure}

To test whether the kinetic shock obeys the MHD scaling relation and evolution, we perform a series of simulations with varying $\sigma_{\rm bg, nl}$ and $\lambda$. We measure $\gamma_{u}$ upstream of the shock where the linear acceleration zone ends, and record the maximum value $\gamma_{u}$ attains over the history as $\gamma_{\rm max}$. The top panel of Figure~\ref{fig:Lorentz} shows that $\gamma_{\rm max}/\lambda$ depends linearly on $\sigma_{\rm bg, nl}$, as predicted by MHD. However, we measured a slightly larger prefactor, so the resulting scaling relation is
\begin{equation}\label{eq:gamma_max}
    \gamma_{\rm max} \approx 
    1.3\,
    \frac{\sigma_{\rm bg,nl}}{2\pi}\left(\frac{\lambda}{r_{\rm nl}}\right).
\end{equation}
In other words, the bulk acceleration is even more efficient than predicted by~\cite{Beloborodov2023}. The bottom panel of Figure~\ref{fig:Lorentz} shows $\gamma_{u}$ as a function of position for three different simulations, normalized by their own maximum value. We see that the evolution of $\gamma_{u}$ in our simulations agrees well with the theoretical prediction, especially during early stages of the shock propagation. At late stages, the measured $\gamma_{u}$ decreases with $r$ slower than the theoretical prediction, closer to $1/r^{3}$. This may depend on the background magnetization $\sigma_{\rm bg,nl}$, as Equation~\eqref{eq:full_monster} was derived for extremely high $\sigma_{\rm bg,nl}$.

\begin{figure*}
    \centering
    \includegraphics[width=0.99\linewidth]{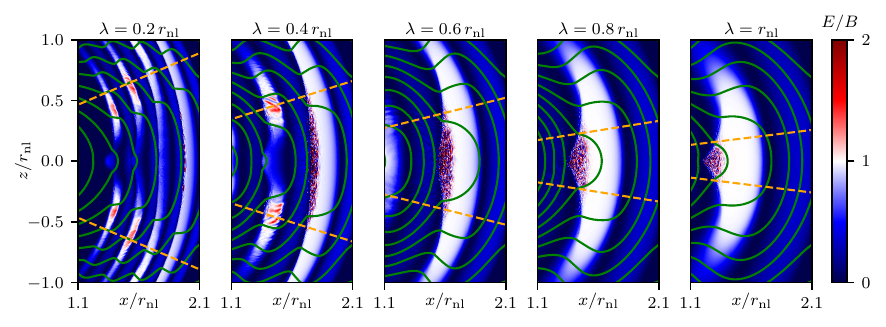}
    \caption{The ratio of the total electric field to the total magnetic field for simulations with several different wavelengths. The fluctuating region with $E>B$ that occurs in the first \textbf{full period of the fast wave} is the precursor wave. The orange dashed lines mark the latitudes where the shock becomes parallel, namely, the magnetic field is parallel to the shock normal. From left to right, these latitudes are $\pm 23^{\circ}$, $\pm 18^{\circ}$, $\pm 14^{\circ}$, $\pm 9^{\circ}$, and $\pm 7^{\circ}$ respectively. Subsequent shocks become modified due to heating of the plasma from the first shock, and $E > B$ regions show up at higher latitudes due to plasma flowing towards the equator. The snapshots are all taken at the same time $t=2r_{\rm nl}/c$.
    }
    \label{fig:shock_front}
\end{figure*}

Away from the equatorial plane, the plasma flow is no longer perpendicular to the shock. The upstream velocity is dominated by the $\mathbf{E} \times \mathbf{B}$ drift, which develops an increasing $\hat{\theta}$ component at higher latitudes, as shown in the rightmost panel of Figure~\ref{fig:Global}. Therefore, the upstream flow at high latitudes is oblique with respect to the shock normal. Interestingly, our simulations demonstrate that the scaling of the maximum upstream Lorentz factor still depends linearly on the background magnetization. This flow also directs plasma from higher latitudes towards the equator, increasing the density in the equatorial plane while decreasing the density at higher latitudes. Consequently the density at high latitudes may decrease enough to no longer prevent $E>B$ in subsequent shocks (see discussion in the Supplemental Materials).

Another important factor that determines the shock properties is the magnetic obliquity, defined as the angle between the magnetic field and the shock normal in the \emph{upstream} rest frame~\cite{2009ApJ...698.1523S}. The orientation of the magnetic field with respect to the shock generally depends on the reference frame. However, when the magnetic field is parallel to the shock normal, an arbitrary Lorentz transformation does not change this angle, therefore we can directly identify a parallel shock in our simulation frame. For all of our simulations, we do see that the shock becomes parallel at sufficiently high latitudes. We measured the latitudes $\pm \ell_{\parallel}$ where the shock becomes parallel, and found that the value does not evolve much over time after the shock has formed; it has a strong dependence only on the fast wave wavelength $\lambda$. The results are shown in Figure~\ref{fig:shock_front}. We found that within the equatorial region bounded by the critical latitudes $\pm \ell_{\parallel}$, the upstream drifts inward toward the shock; at higher latitudes beyond $\pm \ell_{\parallel}$, the upstream drifts outward and the shock is catching up with the upstream flow instead.

Figure~\ref{fig:shock_front} also shows that the shape of the shock front depends on the fast wave wavelength $\lambda$. The shock front can be visualized as the region separating the upstream, where $E \approx B$ from the downstream, where $E/B < 1$. When $\lambda\ll r_{\rm nl}$, the shock front is approximately spherical. However, as $\lambda$ approaches $r_{\rm nl}$, the shock front gets deformed from a spherical shape: the equatorial section lags behind the segments at higher latitudes. The effect is most extreme for the run with $\lambda=r_{\rm nl}$. We discuss the reason behind this in the End Matter.

As an important feature of magnetized shocks, the formation of precursor waves is dependent on the magnetic obliquity. Figure~\ref{fig:shock_front} shows that the region with precursor wave production is around the equator, roughly bounded by the latitudes where the shock becomes parallel. This suggests that precursor waves are most efficiently produced at quasi-perpendicular shocks, when the upstream is accelerated towards the shock. They are suppressed when the shock becomes parallel. We further discuss a few kinematic considerations that affect the appearance of the precursor waves in the End Matter.

\paragraph{Astrophysical Implications.---}
Our global kinetic simulations have shown that the monster shock can indeed emit a coherent precursor wave, and this may be a promising mechanism for FRBs~\cite{2025PhRvL.134c5201V}. For a typical fast wave of luminosity $L = L_{42}10^{42}\, {\rm erg \, s^{-1}}$ and frequency $\omega=\omega_4 10^4\,{\rm rad\,s}^{-1}$ launched from a magnetar of surface magnetic field $B_* = B_{15}10^{15}\, {\rm G}$, we estimate the peak frequency of the radio emission when the shock has just formed beyond the nonlinear radius to be $\nu_{\rm peak}(r_{\rm nl})\sim0.22 \, B_{15}^{1/2}L_{42}^{-1/4}M_6P_0^{-1}\omega_4 \,{\rm GHz}$, and the peak luminosity of the radio waves is $L_R(r_{\rm nl})\sim 5.4\times10^{37}B_{15}^{-1/2}L_{42}^{3/4}\omega_4^{-1}\,{\rm erg\,s}^{-1}$ (see the End Matter for the detailed calculations). Afterwards, $\nu_{\rm peak}(r)$ increases with $r$ linearly, but the luminosity quickly decreases with radius as $L_R\propto r^{-5}$, so most of the radio emission is produced between $r_{\rm nl}<r<3r_{\rm nl}$. The duration of the observed radio emission is $\Delta t_{\rm obs}\sim 0.5\omega_4^{-1}\,{\rm ms}$ if we only consider the first shock, but the subsequent shocks could continue to emit and produce substructures in the observed bursts similar to those seen in some FRBs~\cite{2021ApJ...919L...6M}. We thus conclude that with appropriate parameters, the monster shock is promising to produce observed FRBs, especially the one from the Galactic magnetar SGR 1935+2154. The shock will also produce a multiwavelength counterpart from incoherent radiation in the downstream. This emission is discussed in the end matter.

\paragraph{Conclusion.---}
We have carried out the first 2D global PIC simulations of monster shock formation due to nonlinear steepening of fast waves in neutron star magnetospheres. We confirmed efficient conversion of electromagnetic energy into plasma kinetic energy: the maximum upstream Lorentz factor is indeed linearly proportional to the background magnetization and the fast wave wavelength. We observed the production of precursor waves from the monster shock and measured its angular range. We found that with typical magnetar parameters, the monster shock can be a promising mechanism for some FRBs, especially the ones from the Galactic magnetar SGR 1935+2154. However, more study is needed to investigate whether these radio waves can escape from the magnetosphere. We are also in need of a systematic study of precursor waves emitted from shocks of general magnetic obliquity to obtain more precise predictions of the radio spectra and polarization away from the magnetic equator.

\section{Acknowledgment}
We thank Andrei Beloborodov, Michael Grehan, Amir Levinson, Yuanhong Qu, Andrew Sullivan, and Arno Vanthieghem for helpful discussions. AC and YY acknowledge support from NSF grants DMS-2235457 and AST-2308111. AC also acknowledges support from NASA grant 80NSSC24K1095. This work was also facilitated by the Multimessenger Plasma Physics Center (MPPC), NSF grant PHY-2206608, and by a grant from the Simons Foundation (MP-SCMPS-00001470) to YY. This research used resources of the Oak
Ridge Leadership Computing Facility at the Oak Ridge National Laboratory,
which is supported by the Office of Science of the U.S. Department of Energy
under Contract No. DE-AC05-00OR22725.
We thank the hospitality of the Kavli Institute for Theoretical Physics (KITP), where we had fruitful discussions with colleagues as we revised the manuscript. This research was supported in part by grant NSF PHY-2309135 to the KITP.

% Create the reference section using BibTeX:
\bibliography{ref.bib}

@ARTICLE{2020MNRAS.499.2884B,
       author = {{Babul}, Aliya-Nur and {Sironi}, Lorenzo},
        title = "{The synchrotron maser emission from relativistic magnetized shocks: dependence on the pre-shock temperature}",
      journal = {\mnras},
         year = 2020,
        month = dec,
       volume = {499},
       number = {2},
        pages = {2884-2895},
          doi = {10.1093/mnras/staa2612},
archivePrefix = {arXiv},
       eprint = {2006.03081},
 primaryClass = {astro-ph.HE},
       adsurl = {https://ui.adsabs.harvard.edu/abs/2020MNRAS.499.2884B}
}

@ARTICLE{2021NatAs...5..378L,
       author = {{Li}, C.~K. and {Lin}, L. and {Xiong}, S.~L. and {Ge}, M.~Y. and {Li}, X.~B. and {Li}, T.~P. and {Lu}, F.~J. and {Zhang}, S.~N. and {Tuo}, Y.~L. and {Nang}, Y. and {Zhang}, B. and {Xiao}, S. and {Chen}, Y. and {Song}, L.~M. and {Xu}, Y.~P. and {Liu}, C.~Z. and {Jia}, S.~M. and {Cao}, X.~L. and {Qu}, J.~L. and {Zhang}, S. and {Gu}, Y.~D. and {Liao}, J.~Y. and {Zhao}, X.~F. and {Tan}, Y. and {Nie}, J.~Y. and {Zhao}, H.~S. and {Zheng}, S.~J. and {Zheng}, Y.~G. and {Luo}, Q. and {Cai}, C. and {Li}, B. and {Xue}, W.~C. and {Bu}, Q.~C. and {Chang}, Z. and {Chen}, G. and {Chen}, L. and {Chen}, T.~X. and {Chen}, Y.~B. and {Chen}, Y.~P. and {Cui}, W. and {Cui}, W.~W. and {Deng}, J.~K. and {Dong}, Y.~W. and {Du}, Y.~Y. and {Fu}, M.~X. and {Gao}, G.~H. and {Gao}, H. and {Gao}, M. and {Gu}, Y.~D. and {Guan}, J. and {Guo}, C.~C. and {Han}, D.~W. and {Huang}, Y. and {Huo}, J. and {Jiang}, L.~H. and {Jiang}, W.~C. and {Jin}, J. and {Jin}, Y.~J. and {Kong}, L.~D. and {Li}, G. and {Li}, M.~S. and {Li}, W. and {Li}, X. and {Li}, X.~F. and {Li}, Y.~G. and {Li}, Z.~W. and {Liang}, X.~H. and {Liu}, B.~S. and {Liu}, G.~Q. and {Liu}, H.~W. and {Liu}, X.~J. and {Liu}, Y.~N. and {Lu}, B. and {Lu}, X.~F. and {Luo}, T. and {Ma}, X. and {Meng}, B. and {Ou}, G. and {Sai}, N. and {Shang}, R.~C. and {Song}, X.~Y. and {Sun}, L. and {Tao}, L. and {Wang}, C. and {Wang}, G.~F. and {Wang}, J. and {Wang}, W.~S. and {Wang}, Y.~S. and {Wen}, X.~Y. and {Wu}, B.~B. and {Wu}, B.~Y. and {Wu}, M. and {Xiao}, G.~C. and {Xu}, H. and {Yang}, J.~W. and {Yang}, S. and {Yang}, Y.~J. and {Yang}, Yi-Jung and {Yi}, Q.~B. and {Yin}, Q.~Q. and {You}, Y. and {Zhang}, A.~M. and {Zhang}, C.~M. and {Zhang}, F. and {Zhang}, H.~M. and {Zhang}, J. and {Zhang}, T. and {Zhang}, W. and {Zhang}, W.~C. and {Zhang}, W.~Z. and {Zhang}, Y. and {Zhang}, Yue and {Zhang}, Y.~F. and {Zhang}, Y.~J. and {Zhang}, Z. and {Zhang}, Zhi and {Zhang}, Z.~L. and {Zhou}, D.~K. and {Zhou}, J.~F. and {Zhu}, Y. and {Zhu}, Y.~X. and {Zhuang}, R.~L.},
        title = "{HXMT identification of a non-thermal X-ray burst from SGR J1935+2154 and with FRB 200428}",
      journal = {Nature Astronomy},
         year = 2021,
        month = apr,
       volume = {5},
        pages = {378-384},
          doi = {10.1038/s41550-021-01302-6},
archivePrefix = {arXiv},
       eprint = {2005.11071},
 primaryClass = {astro-ph.HE},
       adsurl = {https://ui.adsabs.harvard.edu/abs/2021NatAs...5..378L}
}

@ARTICLE{2014ApJ...795L..22C,
       author = {{Chen}, Alexander Y. and {Beloborodov}, Andrei M.},
        title = "{Electrodynamics of Axisymmetric Pulsar Magnetosphere with Electron-Positron Discharge: A Numerical Experiment}",
      journal = {\apjl},
         year = 2014,
        month = nov,
       volume = {795},
       number = {1},
          eid = {L22},
        pages = {L22},
          doi = {10.1088/2041-8205/795/1/L22},
archivePrefix = {arXiv},
       eprint = {1406.7834},
 primaryClass = {astro-ph.HE},
       adsurl = {https://ui.adsabs.harvard.edu/abs/2014ApJ...795L..22C}
}

@ARTICLE{Beloborodov2023,
       author = {{Beloborodov}, Andrei M.},
        title = "{Monster Radiative Shocks in the Perturbed Magnetospheres of Neutron Stars}",
      journal = {\apj},
         year = 2023,
        month = dec,
       volume = {959},
       number = {1},
          eid = {34},
        pages = {34},
          doi = {10.3847/1538-4357/acf659},
archivePrefix = {arXiv},
       eprint = {2210.13509},
 primaryClass = {astro-ph.HE},
       adsurl = {https://ui.adsabs.harvard.edu/abs/2023ApJ...959...34B}
}

@ARTICLE{2009ApJ...698.1523S,
       author = {{Sironi}, Lorenzo and {Spitkovsky}, Anatoly},
        title = "{Particle Acceleration in Relativistic Magnetized Collisionless Pair Shocks: Dependence of Shock Acceleration on Magnetic Obliquity}",
      journal = {\apj},
         year = 2009,
        month = jun,
       volume = {698},
       number = {2},
        pages = {1523-1549},
          doi = {10.1088/0004-637X/698/2/1523},
archivePrefix = {arXiv},
       eprint = {0901.2578},
 primaryClass = {astro-ph.HE},
       adsurl = {https://ui.adsabs.harvard.edu/abs/2009ApJ...698.1523S}
}

@ARTICLE{Plotnikov2019,
       author = {{Plotnikov}, Illya and {Sironi}, Lorenzo},
        title = "{The synchrotron maser emission from relativistic shocks in Fast Radio Bursts: 1D PIC simulations of cold pair plasmas}",
      journal = {\mnras},
         year = 2019,
        month = may,
       volume = {485},
       number = {3},
        pages = {3816-3833},
          doi = {10.1093/mnras/stz640},
archivePrefix = {arXiv},
       eprint = {1901.01029},
 primaryClass = {astro-ph.HE},
       adsurl = {https://ui.adsabs.harvard.edu/abs/2019MNRAS.485.3816P}
}

@ARTICLE{2022arXiv221013506C,
       author = {{Chen}, Alexander Y. and {Yuan}, Yajie and {Li}, Xinyu and {Mahlmann}, Jens F.},
        title = "{Propagation of a Strong Fast Magnetosonic Wave in the Magnetosphere of a Neutron Star}",
      journal = {arXiv e-prints},
         year = 2022,
        month = oct,
          eid = {arXiv:2210.13506},
        pages = {arXiv:2210.13506},
          doi = {10.48550/arXiv.2210.13506},
archivePrefix = {arXiv},
       eprint = {2210.13506},
 primaryClass = {astro-ph.HE},
       adsurl = {https://ui.adsabs.harvard.edu/abs/2022arXiv221013506C}
}

@ARTICLE{2025PhRvE.111d5209V,
       author = {{Vanthieghem}, Arno and {Levinson}, Amir},
        title = "{Relativistically magnetized collisionless shocks in pair plasma: Solitons, chaos, and thermalization}",
      journal = {\pre},
         year = 2025,
        month = apr,
       volume = {111},
       number = {4},
          eid = {045209},
        pages = {045209},
          doi = {10.1103/PhysRevE.111.045209},
archivePrefix = {arXiv},
       eprint = {2411.16484},
 primaryClass = {astro-ph.HE},
       adsurl = {https://ui.adsabs.harvard.edu/abs/2025PhRvE.111d5209V}
}

@ARTICLE{2025PhRvL.134c5201V,
       author = {{Vanthieghem}, A. and {Levinson}, A.},
        title = "{Fast Radio Bursts as Precursor Radio Emission from Monster Shocks}",
      journal = {\prl},
         year = 2025,
        month = jan,
       volume = {134},
       number = {3},
          eid = {035201},
        pages = {035201},
          doi = {10.1103/PhysRevLett.134.035201},
archivePrefix = {arXiv},
       eprint = {2407.15076},
 primaryClass = {astro-ph.HE},
       adsurl = {https://ui.adsabs.harvard.edu/abs/2025PhRvL.134c5201V}
}

@ARTICLE{2003MNRAS.339..765L,
       author = {{Lyubarsky}, Y.~E.},
        title = "{Fast magnetosonic waves in pulsar winds}",
      journal = {\mnras},
         year = 2003,
        month = mar,
       volume = {339},
       number = {3},
        pages = {765-771},
          doi = {10.1046/j.1365-8711.2003.06221.x},
archivePrefix = {arXiv},
       eprint = {astro-ph/0211046},
 primaryClass = {astro-ph},
       adsurl = {https://ui.adsabs.harvard.edu/abs/2003MNRAS.339..765L}
}

@ARTICLE{2024ApJ...974L..12M,
       author = {{Most}, Elias R. and {Beloborodov}, Andrei M. and {Ripperda}, Bart},
        title = "{Monster Shocks, Gamma-Ray Bursts, and Black Hole Quasi-normal Modes from Neutron-star Collapse}",
      journal = {\apjl},
         year = 2024,
        month = oct,
       volume = {974},
       number = {1},
          eid = {L12},
        pages = {L12},
          doi = {10.3847/2041-8213/ad7e1f},
archivePrefix = {arXiv},
       eprint = {2404.01456},
 primaryClass = {astro-ph.HE},
       adsurl = {https://ui.adsabs.harvard.edu/abs/2024ApJ...974L..12M}
}

@ARTICLE{2025ApJ...982L..54K,
       author = {{Kim}, Yoonsoo and {Most}, Elias R. and {Beloborodov}, Andrei M. and {Ripperda}, Bart},
        title = "{Black Hole Pulsars and Monster Shocks as Outcomes of Black Hole{\textendash}Neutron Star Mergers}",
      journal = {\apjl},
         year = 2025,
        month = apr,
       volume = {982},
       number = {2},
          eid = {L54},
        pages = {L54},
          doi = {10.3847/2041-8213/adbff9},
archivePrefix = {arXiv},
       eprint = {2412.05760},
 primaryClass = {astro-ph.HE},
       adsurl = {https://ui.adsabs.harvard.edu/abs/2025ApJ...982L..54K}
}

@ARTICLE{2025arXiv250304558C,
       author = {{Chen}, Alexander Y. and {Luepker}, Martin and {Yuan}, Yajie},
        title = "{Introducing APERTURE: A GPU-based General Relativistic Particle-in-Cell Simulation Framework}",
      journal = {arXiv e-prints},
         year = 2025,
        month = mar,
          eid = {arXiv:2503.04558},
        pages = {arXiv:2503.04558},
          doi = {10.48550/arXiv.2503.04558},
archivePrefix = {arXiv},
       eprint = {2503.04558},
 primaryClass = {astro-ph.HE},
       adsurl = {https://ui.adsabs.harvard.edu/abs/2025arXiv250304558C}
}

@ARTICLE{2001CoPhC.135..144E,
       author = {{Esirkepov}, T. Zh.},
        title = "{Exact charge conservation scheme for Particle-in-Cell simulation with an arbitrary form-factor}",
      journal = {Computer Physics Communications},
         year = 2001,
        month = apr,
       volume = {135},
       number = {2},
        pages = {144-153},
          doi = {10.1016/S0010-4655(00)00228-9},
       adsurl = {https://ui.adsabs.harvard.edu/abs/2001CoPhC.135..144E}
}

@ARTICLE{2008PhPl...15e6701V,
       author = {{Vay}, J. -L.},
        title = "{Simulation of beams or plasmas crossing at relativistic velocity}",
      journal = {Physics of Plasmas},
         year = 2008,
        month = may,
       volume = {15},
       number = {5},
          eid = {056701},
        pages = {056701},
          doi = {10.1063/1.2837054},
       adsurl = {https://ui.adsabs.harvard.edu/abs/2008PhPl...15e6701V}
}

@ARTICLE{2025arXiv250812567Q,
       author = {{Qu}, Yuanhong and {Bransgrove}, Ashley},
        title = "{Three-Dimensional Numerical Simulations of Magnetar Crust Quakes}",
      journal = {arXiv e-prints},
         year = 2025,
        month = aug,
          eid = {arXiv:2508.12567},
        pages = {arXiv:2508.12567},
          doi = {10.48550/arXiv.2508.12567},
archivePrefix = {arXiv},
       eprint = {2508.12567},
 primaryClass = {astro-ph.HE},
       adsurl = {https://ui.adsabs.harvard.edu/abs/2025arXiv250812567Q}
}

@ARTICLE{2025arXiv250818033B,
       author = {{Burnaz}, Louis and {Most}, Elias R. and {Bransgrove}, Ashley},
        title = "{Crustal Quakes Spark Magnetospheric Blasts: Imprints of Realistic Magnetar Crust Oscillations on the Fast Radio Burst Signal}",
      journal = {arXiv e-prints},
         year = 2025,
        month = aug,
          eid = {arXiv:2508.18033},
        pages = {arXiv:2508.18033},
          doi = {10.48550/arXiv.2508.18033},
archivePrefix = {arXiv},
       eprint = {2508.18033},
 primaryClass = {astro-ph.HE},
       adsurl = {https://ui.adsabs.harvard.edu/abs/2025arXiv250818033B}
}

@ARTICLE{2020ApJ...898L..29M,
       author = {{Mereghetti}, S. and {Savchenko}, V. and {Ferrigno}, C. and {G{\"o}tz}, D. and {Rigoselli}, M. and {Tiengo}, A. and {Bazzano}, A. and {Bozzo}, E. and {Coleiro}, A. and {Courvoisier}, T.~J. -L. and {Doyle}, M. and {Goldwurm}, A. and {Hanlon}, L. and {Jourdain}, E. and {von Kienlin}, A. and {Lutovinov}, A. and {Martin-Carrillo}, A. and {Molkov}, S. and {Natalucci}, L. and {Onori}, F. and {Panessa}, F. and {Rodi}, J. and {Rodriguez}, J. and {S{\'a}nchez-Fern{\'a}ndez}, C. and {Sunyaev}, R. and {Ubertini}, P.},
        title = "{INTEGRAL Discovery of a Burst with Associated Radio Emission from the Magnetar SGR 1935+2154}",
      journal = {\apjl},
         year = 2020,
        month = aug,
       volume = {898},
       number = {2},
          eid = {L29},
        pages = {L29},
          doi = {10.3847/2041-8213/aba2cf},
archivePrefix = {arXiv},
       eprint = {2005.06335},
 primaryClass = {astro-ph.HE},
       adsurl = {https://ui.adsabs.harvard.edu/abs/2020ApJ...898L..29M}
}

@ARTICLE{2021NatAs...5..372R,
       author = {{Ridnaia}, A. and {Svinkin}, D. and {Frederiks}, D. and {Bykov}, A. and {Popov}, S. and {Aptekar}, R. and {Golenetskii}, S. and {Lysenko}, A. and {Tsvetkova}, A. and {Ulanov}, M. and {Cline}, T.~L.},
        title = "{A peculiar hard X-ray counterpart of a Galactic fast radio burst}",
      journal = {Nature Astronomy},
         year = 2021,
        month = apr,
       volume = {5},
        pages = {372-377},
          doi = {10.1038/s41550-020-01265-0},
archivePrefix = {arXiv},
       eprint = {2005.11178},
 primaryClass = {astro-ph.HE},
       adsurl = {https://ui.adsabs.harvard.edu/abs/2021NatAs...5..372R}
}

@ARTICLE{Zhang_2025,
       author = {{Zhang}, Yu and {Yang}, Yuanpei and {Ji}, Liangliang},
        title = "{Radiation Reaction effects on Coherent Emission in Relativistic Magnetized Shocks}",
      journal = {arXiv e-prints},
         year = 2025,
        month = feb,
          eid = {arXiv:2502.14550},
        pages = {arXiv:2502.14550},
archivePrefix = {arXiv},
 primaryClass = {astro-ph.HE},
       adsurl = {https://ui.adsabs.harvard.edu/abs/2025arXiv250214550Z}
}

@ARTICLE{1989ApJ...343..839B,
       author = {{Blaes}, O. and {Blandford}, R. and {Goldreich}, P. and {Madau}, P.},
        title = "{Neutron Starquake Models for Gamma-Ray Bursts}",
      journal = {\apj},
         year = 1989,
        month = aug,
       volume = {343},
        pages = {839},
          doi = {10.1086/167754},
       adsurl = {https://ui.adsabs.harvard.edu/abs/1989ApJ...343..839B}
}

@ARTICLE{1969ApJ...157..869G,
       author = {{Goldreich}, Peter and {Julian}, William H.},
        title = "{Pulsar Electrodynamics}",
      journal = {\apj},
         year = 1969,
        month = aug,
       volume = {157},
        pages = {869},
          doi = {10.1086/150119},
       adsurl = {https://ui.adsabs.harvard.edu/abs/1969ApJ...157..869G}
}

@ARTICLE{1995MNRAS.275..255T,
       author = {{Thompson}, Christopher and {Duncan}, Robert C.},
        title = "{The soft gamma repeaters as very strongly magnetized neutron stars - I. Radiative mechanism for outbursts}",
      journal = {\mnras},
         year = 1995,
        month = jul,
       volume = {275},
       number = {2},
        pages = {255-300},
          doi = {10.1093/mnras/275.2.255},
       adsurl = {https://ui.adsabs.harvard.edu/abs/1995MNRAS.275..255T}
}

@ARTICLE{1996ApJ...473..322T,
       author = {{Thompson}, Christopher and {Duncan}, Robert C.},
        title = "{The Soft Gamma Repeaters as Very Strongly Magnetized Neutron Stars. II. Quiescent Neutrino, X-Ray, and Alfven Wave Emission}",
      journal = {\apj},
         year = 1996,
        month = dec,
       volume = {473},
        pages = {322},
          doi = {10.1086/178147},
       adsurl = {https://ui.adsabs.harvard.edu/abs/1996ApJ...473..322T}
}

@ARTICLE{1988PhFl...31..839A,
       author = {{Alsop}, David and {Arons}, Jonathan},
        title = "{Relativistic magnetosonic solitons with reflected particles in electron-positron plasmas}",
      journal = {Physics of Fluids},
         year = 1988,
        month = apr,
       volume = {31},
       number = {4},
        pages = {839-847},
          doi = {10.1063/1.866765},
       adsurl = {https://ui.adsabs.harvard.edu/abs/1988PhFl...31..839A}
}

@ARTICLE{1992ApJ...391...73G,
       author = {{Gallant}, Yves A. and {Hoshino}, Masahiro and {Langdon}, A.~B. and {Arons}, Jonathan and {Max}, Claire E.},
        title = "{Relativistic, Perpendicular Shocks in Electron-Positron Plasmas}",
      journal = {\apj},
         year = 1992,
        month = may,
       volume = {391},
        pages = {73},
          doi = {10.1086/171326},
       adsurl = {https://ui.adsabs.harvard.edu/abs/1992ApJ...391...73G}
}

@ARTICLE{2025ApJ...987...42C,
       author = {{Chen}, Alexander Y. and {Yuan}, Yajie and {Bernardi}, Dominic},
        title = "{Alfv{\'e}n Wave Mode Conversion in Neutron Star Magnetospheres: A Semianalytic Approach}",
      journal = {\apj},
         year = 2025,
        month = jul,
       volume = {987},
       number = {1},
          eid = {42},
        pages = {42},
          doi = {10.3847/1538-4357/adda3b},
archivePrefix = {arXiv},
       eprint = {2404.06431},
 primaryClass = {astro-ph.HE},
       adsurl = {https://ui.adsabs.harvard.edu/abs/2025ApJ...987...42C}
}

@ARTICLE{2022Natur.607..256C,
       author = {{Chime/Frb Collaboration}, Bridget C., Andersen and {Bandura}, Kevin and {Bhardwaj}, Mohit and {Boyle}, P.~J. and {Brar}, Charanjot and {Breitman}, Daniela and {Cassanelli}, Tomas and {Chatterjee}, Shami and {Chawla}, Pragya and {Cliche}, Jean-Fran{\c{c}}ois and {Cubranic}, Davor and {Curtin}, Alice P. and {Deng}, Meiling and {Dobbs}, Matt and {Dong}, Fengqiu Adam and {Fonseca}, Emmanuel and {Gaensler}, B.~M. and {Giri}, Utkarsh and {Good}, Deborah C. and {Hill}, Alex S. and {Josephy}, Alexander and {Kaczmarek}, J.~F. and {Kader}, Zarif and {Kania}, Joseph and {Kaspi}, Victoria M. and {Leung}, Calvin and {Li}, D.~Z. and {Lin}, Hsiu-Hsien and {Masui}, Kiyoshi W. and {McKinven}, Ryan and {Mena-Parra}, Juan and {Merryfield}, Marcus and {Meyers}, B.~W. and {Michilli}, D. and {Naidu}, Arun and {Newburgh}, Laura and {Ng}, C. and {Ordog}, Anna and {Patel}, Chitrang and {Pearlman}, Aaron B. and {Pen}, Ue-Li and {Petroff}, Emily and {Pleunis}, Ziggy and {Rafiei-Ravandi}, Masoud and {Rahman}, Mubdi and {Ransom}, Scott and {Renard}, Andre and {Sanghavi}, Pranav and {Scholz}, Paul and {Shaw}, J. Richard and {Shin}, Kaitlyn and {Siegel}, Seth R. and {Singh}, Saurabh and {Smith}, Kendrick and {Stairs}, Ingrid and {Tan}, Chia Min and {Tendulkar}, Shriharsh P. and {Vanderlinde}, Keith and {Wiebe}, D.~V. and {Wulf}, Dallas and {Zwaniga}, Andrew},
        title = "{Sub-second periodicity in a fast radio burst}",
      journal = {\nat},
         year = 2022,
        month = jul,
       volume = {607},
       number = {7918},
        pages = {256-259},
          doi = {10.1038/s41586-022-04841-8},
archivePrefix = {arXiv},
       eprint = {2107.08463},
 primaryClass = {astro-ph.HE},
       adsurl = {https://ui.adsabs.harvard.edu/abs/2022Natur.607..256C}
}

@ARTICLE{2021ApJ...919L...6M,
       author = {{Majid}, Walid A. and {Pearlman}, Aaron B. and {Prince}, Thomas A. and {Wharton}, Robert S. and {Naudet}, Charles J. and {Bansal}, Karishma and {Connor}, Liam and {Bhardwaj}, Mohit and {Tendulkar}, Shriharsh P.},
        title = "{A Bright Fast Radio Burst from FRB 20200120E with Sub-100 Nanosecond Structure}",
      journal = {\apjl},
         year = 2021,
        month = sep,
       volume = {919},
       number = {1},
          eid = {L6},
        pages = {L6},
          doi = {10.3847/2041-8213/ac1921},
archivePrefix = {arXiv},
       eprint = {2105.10987},
 primaryClass = {astro-ph.HE},
       adsurl = {https://ui.adsabs.harvard.edu/abs/2021ApJ...919L...6M}
}

@phdthesis{chen-thesis,
  author       = {{Chen}, A.~Y.},
  title        = {Particle-in-Cell Simulations and their Applications to Magnetospheres of Neutron Stars},
  school       = {Columbia University},
  year         = 2017,
  month        = 8,
  doi          = {10.7916/D80V8R6G},
  url          = {http://dx.doi.org/10.7916/D80V8R6G}
}

@ARTICLE{2020Natur.587...54C,
       author = {{CHIME/FRB Collaboration} and {Andersen}, B.~C. and {Bandura}, K.~M. and {Bhardwaj}, M. and {Bij}, A. and {Boyce}, M.~M. and {Boyle}, P.~J. and {Brar}, C. and {Cassanelli}, T. and {Chawla}, P. and {Chen}, T. and {Cliche}, J. -F. and {Cook}, A. and {Cubranic}, D. and {Curtin}, A.~P. and {Denman}, N.~T. and {Dobbs}, M. and {Dong}, F.~Q. and {Fandino}, M. and {Fonseca}, E. and {Gaensler}, B.~M. and {Giri}, U. and {Good}, D.~C. and {Halpern}, M. and {Hill}, A.~S. and {Hinshaw}, G.~F. and {H{\"o}fer}, C. and {Josephy}, A. and {Kania}, J.~W. and {Kaspi}, V.~M. and {Landecker}, T.~L. and {Leung}, C. and {Li}, D.~Z. and {Lin}, H. -H. and {Masui}, K.~W. and {McKinven}, R. and {Mena-Parra}, J. and {Merryfield}, M. and {Meyers}, B.~W. and {Michilli}, D. and {Milutinovic}, N. and {Mirhosseini}, A. and {M{\"u}nchmeyer}, M. and {Naidu}, A. and {Newburgh}, L.~B. and {Ng}, C. and {Patel}, C. and {Pen}, U. -L. and {Pinsonneault-Marotte}, T. and {Pleunis}, Z. and {Quine}, B.~M. and {Rafiei-Ravandi}, M. and {Rahman}, M. and {Ransom}, S.~M. and {Renard}, A. and {Sanghavi}, P. and {Scholz}, P. and {Shaw}, J.~R. and {Shin}, K. and {Siegel}, S.~R. and {Singh}, S. and {Smegal}, R.~J. and {Smith}, K.~M. and {Stairs}, I.~H. and {Tan}, C.~M. and {Tendulkar}, S.~P. and {Tretyakov}, I. and {Vanderlinde}, K. and {Wang}, H. and {Wulf}, D. and {Zwaniga}, A.~V.},
        title = "{A bright millisecond-duration radio burst from a Galactic magnetar}",
      journal = {\nat},
         year = 2020,
        month = nov,
       volume = {587},
       number = {7832},
        pages = {54-58},
          doi = {10.1038/s41586-020-2863-y},
archivePrefix = {arXiv},
       eprint = {2005.10324},
 primaryClass = {astro-ph.HE},
       adsurl = {https://ui.adsabs.harvard.edu/abs/2020Natur.587...54C}
}

@ARTICLE{2020Natur.587...59B,
       author = {{Bochenek}, C.~D. and {Ravi}, V. and {Belov}, K.~V. and {Hallinan}, G. and {Kocz}, J. and {Kulkarni}, S.~R. and {McKenna}, D.~L.},
        title = "{A fast radio burst associated with a Galactic magnetar}",
      journal = {\nat},
         year = 2020,
        month = nov,
       volume = {587},
       number = {7832},
        pages = {59-62},
          doi = {10.1038/s41586-020-2872-x},
archivePrefix = {arXiv},
       eprint = {2005.10828},
 primaryClass = {astro-ph.HE},
       adsurl = {https://ui.adsabs.harvard.edu/abs/2020Natur.587...59B}
}

@ARTICLE{2017ARA&A..55..261K,
       author = {{Kaspi}, Victoria M. and {Beloborodov}, Andrei M.},
        title = "{Magnetars}",
      journal = {\araa},
         year = 2017,
        month = aug,
       volume = {55},
       number = {1},
        pages = {261-301},
          doi = {10.1146/annurev-astro-081915-023329},
archivePrefix = {arXiv},
       eprint = {1703.00068},
 primaryClass = {astro-ph.HE},
       adsurl = {https://ui.adsabs.harvard.edu/abs/2017ARA&A..55..261K}
}

@ARTICLE{2020ApJ...897..173B,
       author = {{Bransgrove}, Ashley and {Beloborodov}, Andrei M. and {Levin}, Yuri},
        title = "{A Quake Quenching the Vela Pulsar}",
      journal = {\apj},
         year = 2020,
        month = jul,
       volume = {897},
       number = {2},
          eid = {173},
        pages = {173},
          doi = {10.3847/1538-4357/ab93b7},
archivePrefix = {arXiv},
       eprint = {2001.08658},
 primaryClass = {astro-ph.HE},
       adsurl = {https://ui.adsabs.harvard.edu/abs/2020ApJ...897..173B}
}

@ARTICLE{2015SSRv..191..315M,
       author = {{Mereghetti}, Sandro and {Pons}, Jos{\'e} A. and {Melatos}, Andrew},
        title = "{Magnetars: Properties, Origin and Evolution}",
      journal = {\ssr},
         year = 2015,
        month = oct,
       volume = {191},
       number = {1-4},
        pages = {315-338},
          doi = {10.1007/s11214-015-0146-y},
archivePrefix = {arXiv},
       eprint = {1503.06313},
 primaryClass = {astro-ph.HE},
       adsurl = {https://ui.adsabs.harvard.edu/abs/2015SSRv..191..315M}
}

@ARTICLE{2024ApJ...972..139M,
       author = {{Mahlmann}, Jens F. and {Aloy}, Miguel {\'A}. and {Li}, Xinyu},
        title = "{Force-free Wave Interaction in Magnetar Magnetospheres: Computational Modeling in Axisymmetry}",
      journal = {\apj},
         year = 2024,
        month = sep,
       volume = {972},
       number = {2},
          eid = {139},
        pages = {139},
          doi = {10.3847/1538-4357/ad60c4},
archivePrefix = {arXiv},
       eprint = {2405.12272},
 primaryClass = {astro-ph.HE},
       adsurl = {https://ui.adsabs.harvard.edu/abs/2024ApJ...972..139M}
}

@ARTICLE{2025ApJ...980..222B,
       author = {{Bernardi}, Dominic and {Yuan}, Yajie and {Chen}, Alexander Y.},
        title = "{Alfv{\'e}n Wave Conversion to Low Frequency Fast Magnetosonic Waves in Magnetar Magnetospheres}",
      journal = {\apj},
         year = 2025,
        month = feb,
       volume = {980},
       number = {2},
          eid = {222},
        pages = {222},
          doi = {10.3847/1538-4357/adabe5},
archivePrefix = {arXiv},
       eprint = {2405.02199},
 primaryClass = {astro-ph.HE},
       adsurl = {https://ui.adsabs.harvard.edu/abs/2025ApJ...980..222B}
}

@ARTICLE{2025ApJ...991...98S,
       author = {{Sullivan}, Andrew G. and {Cort{\'e}s}, Jorge and {Sironi}, Lorenzo},
        title = "{Polarized Emission of Intrabinary Shocks in Spider Pulsars from Global 3D Kinetic Simulations}",
      journal = {\apj},
         year = 2025,
        month = sep,
       volume = {991},
       number = {1},
          eid = {98},
        pages = {98},
          doi = {10.3847/1538-4357/adfdcd},
archivePrefix = {arXiv},
       eprint = {2508.11625},
 primaryClass = {astro-ph.HE},
       adsurl = {https://ui.adsabs.harvard.edu/abs/2025ApJ...991...98S}
}

@ARTICLE{2018ApJ...855...94P,
       author = {{Philippov}, Alexander A. and {Spitkovsky}, Anatoly},
        title = "{Ab-initio Pulsar Magnetosphere: Particle Acceleration in Oblique Rotators and High-energy Emission Modeling}",
      journal = {\apj},
         year = 2018,
        month = mar,
       volume = {855},
       number = {2},
          eid = {94},
        pages = {94},
          doi = {10.3847/1538-4357/aaabbc},
archivePrefix = {arXiv},
       eprint = {1707.04323},
 primaryClass = {astro-ph.HE},
       adsurl = {https://ui.adsabs.harvard.edu/abs/2018ApJ...855...94P}
}

@ARTICLE{2020A&A...642A.123C,
       author = {{Cerutti}, Beno{\^\i}t and {Giacinti}, Gwenael},
        title = "{A global model of particle acceleration at pulsar wind termination shocks}",
      journal = {\aap},
         year = 2020,
        month = oct,
       volume = {642},
          eid = {A123},
        pages = {A123},
          doi = {10.1051/0004-6361/202038883},
archivePrefix = {arXiv},
       eprint = {2008.07253},
 primaryClass = {astro-ph.HE},
       adsurl = {https://ui.adsabs.harvard.edu/abs/2020A&A...642A.123C}
}

@ARTICLE{2021ApJ...908..176Y,
       author = {{Yuan}, Yajie and {Levin}, Yuri and {Bransgrove}, Ashley and {Philippov}, Alexander},
        title = "{Alfv{\'e}n Wave Mode Conversion in Pulsar Magnetospheres}",
      journal = {\apj},
         year = 2021,
        month = feb,
       volume = {908},
       number = {2},
          eid = {176},
        pages = {176},
          doi = {10.3847/1538-4357/abd405},
archivePrefix = {arXiv},
       eprint = {2007.11504},
 primaryClass = {astro-ph.HE},
       adsurl = {https://ui.adsabs.harvard.edu/abs/2021ApJ...908..176Y}
}

@ARTICLE{2023Sci...382..294R,
       author = {{Ryder}, S.~D. and {Bannister}, K.~W. and {Bhandari}, S. and {Deller}, A.~T. and {Ekers}, R.~D. and {Glowacki}, M. and {Gordon}, A.~C. and {Gourdji}, K. and {James}, C.~W. and {Kilpatrick}, C.~D. and {Lu}, W. and {Marnoch}, L. and {Moss}, V.~A. and {Prochaska}, J.~X. and {Qiu}, H. and {Sadler}, E.~M. and {Simha}, S. and {Sammons}, M.~W. and {Scott}, D.~R. and {Tejos}, N. and {Shannon}, R.~M.},
        title = "{A luminous fast radio burst that probes the Universe at redshift 1}",
      journal = {Science},
         year = 2023,
        month = oct,
       volume = {382},
       number = {6668},
        pages = {294-299},
          doi = {10.1126/science.adf2678},
archivePrefix = {arXiv},
       eprint = {2210.04680},
 primaryClass = {astro-ph.HE},
       adsurl = {https://ui.adsabs.harvard.edu/abs/2023Sci...382..294R}
}

@misc{jackson1999classical,
  title={Classical electrodynamics},
  author={Jackson, John David},
  year={1999},
  publisher={American Association of Physics Teachers}
}

@ARTICLE{2015MNRAS.448..606C,
       author = {{Cerutti}, Beno{\^\i}t and {Philippov}, Alexander and {Parfrey}, Kyle and {Spitkovsky}, Anatoly},
        title = "{Particle acceleration in axisymmetric pulsar current sheets}",
      journal = {\mnras},
         year = 2015,
        month = mar,
       volume = {448},
       number = {1},
        pages = {606-619},
          doi = {10.1093/mnras/stv042},
archivePrefix = {arXiv},
       eprint = {1410.3757},
 primaryClass = {astro-ph.HE},
       adsurl = {https://ui.adsabs.harvard.edu/abs/2015MNRAS.448..606C}
}

@ARTICLE{2017ApJ...844..133C,
       author = {{Chen}, Alexander Y. and {Beloborodov}, Andrei M.},
        title = "{Particle-in-Cell Simulations of the Twisted Magnetospheres of Magnetars. I.}",
      journal = {\apj},
         year = 2017,
        month = aug,
       volume = {844},
       number = {2},
          eid = {133},
        pages = {133},
          doi = {10.3847/1538-4357/aa7a57},
archivePrefix = {arXiv},
       eprint = {1610.10036},
 primaryClass = {astro-ph.HE},
       adsurl = {https://ui.adsabs.harvard.edu/abs/2017ApJ...844..133C}
}

@ARTICLE{2020ApJ...889...69C,
       author = {{Chen}, Alexander Y. and {Cruz}, F{\'a}bio and {Spitkovsky}, Anatoly},
        title = "{Filling the Magnetospheres of Weak Pulsars}",
      journal = {\apj},
         year = 2020,
        month = jan,
       volume = {889},
       number = {1},
          eid = {69},
        pages = {69},
          doi = {10.3847/1538-4357/ab5c20},
archivePrefix = {arXiv},
       eprint = {1911.00059},
 primaryClass = {astro-ph.HE},
       adsurl = {https://ui.adsabs.harvard.edu/abs/2020ApJ...889...69C}
}

@ARTICLE{2022ApJ...933..174Y,
       author = {{Yuan}, Yajie and {Beloborodov}, Andrei M. and {Chen}, Alexander Y. and {Levin}, Yuri and {Most}, Elias R. and {Philippov}, Alexander A.},
        title = "{Magnetar Bursts Due to Alfv{\'e}n Wave Nonlinear Breakout}",
      journal = {\apj},
         year = 2022,
        month = jul,
       volume = {933},
       number = {2},
          eid = {174},
        pages = {174},
          doi = {10.3847/1538-4357/ac7529},
archivePrefix = {arXiv},
       eprint = {2204.08513},
 primaryClass = {astro-ph.HE},
       adsurl = {https://ui.adsabs.harvard.edu/abs/2022ApJ...933..174Y}
}

\clearpage

\section{End Matter}

\paragraph{2D shock structure.---}
We can understand the trend of the shock shape shown in Figure~\ref{fig:shock_front} based on the following considerations. As has been shown by~\cite{Beloborodov2023}, the fast wave becomes nonlinear and forms a shock first on the equator. Then, the higher latitude portion of the wave forms a shock at larger radii. The shape of the shock front is then determined by both the location where it forms, as well as its speed. The latter depends on the local upstream magnetization. In our simulations, the upstream magnetization on the equator can be expressed using the lab-frame-measured magnetic field $B$ and density $n$ as $\sigma_u=B^2/(4\pi n \gamma m_ec^2)$. Just beyond the nonlinear radius, we have $B\sim B_{\rm bg}/2$, $n\sim n_{\rm bg}/2$, and $\gamma\sim\gamma_{\rm max}\sim 1.3(\sigma_{\rm bg,nl}/2\pi)(\lambda/r_{\rm nl})$ according to Equation~\eqref{eq:gamma_max}, so
\begin{equation}\label{eq:sigma_u}
    \sigma_u\sim \frac{\sigma_{\rm bg,nl}}{2\gamma_{\rm max}}\sim\frac{\pi r_{\rm nl}}{1.3\lambda}.
\end{equation}
Interestingly, this does not depend on $\sigma_{\rm bg}$. It is known from MHD theory that when $\sigma_u\gg1$, the shock Lorentz factor in the downstream frame is $\gamma_{\rm sh|d}\sim\sqrt{\sigma_u}$. In the lab frame, the plasma in the downstream is moving nonrelativistically, so the shock speed in the lab frame is $\gamma_{\rm sh}\sim \gamma_{\rm sh|d}\sim\sqrt{\sigma_u}$ when $\sigma_u\gg1$. We can immediately see that $\gamma_{\rm sh}$ decreases with increasing fast wave wavelength $\lambda$. As $\lambda$ approaches $r_{\rm nl}$, the upstream magnetization $\sigma_u$ drops towards order unity, and the shock speed becomes significantly less than the speed of light. Therefore, the equatorial shock propagates slower than the unshocked wave at higher latitudes. When the latter also forms a shock at a larger radius, the equatorial shock is already lagging behind, producing the concave shock shape seen in the right panels of Figure~\ref{fig:shock_front}.

The reason why the shock becomes parallel at smaller latitudes as $\lambda$ decreases is now twofold. Firstly, the concave shock shape makes the shock normal point toward the equator at small latitudes. Secondly, the upstream plasma drifts toward the equator, bringing in the background magnetic field with it---an effect that becomes more prominent when the upstream magnetization is reduced and the plasma inertia becomes significant. In the other limit when $\lambda\ll r_{\rm nl}$, $\sigma_u\gg1$ is satisfied and the shock propagates at a highly relativistic speed. The equatorial shock does not significantly lag behind the unshocked wave at higher latitudes, and the resulting shock surface approaches the spherical limit. In this case, the deformation of the upstream magnetic field due to plasma drift is also negligible, therefore the shock becomes parallel when the upstream magnetic field is purely radial, at a polar angle of $\theta = \sin^{-1}\left(2/\sqrt{5}\right)\approx63^{\circ}$~\cite{Beloborodov2023}, or a latitude of $\ell\approx 27^{\circ}$. This is independent of the angular profile of the original fast wave.

There are a few kinematic considerations that affect the appearance of the precursor waves in this global setting. The precursor waves are likely emitted from the shock surface with a range of wave vector $\mathbf{k}$. However, only the waves emitted within an opening angle of $\sim 1/\gamma_{\rm sh}$ around the local shock normal may outrun the shock. For relatively slowly moving shocks at large fast wave wavelength $\lambda$, this opening angle can be of order unity, therefore the precursor waves may be able to extend beyond the $\ell_\parallel$ bound. In addition, the length of the precursor wave train in the lab frame depends on the Lorentz factor of the shock $\gamma_{\rm sh}$. If the duration of the precursor wave emission is $\Delta t$, then the wave train length in the lab frame is $\Delta l\sim c\Delta t/\gamma_{\rm sh}^2$ in the regime $\gamma_{\rm sh}\gg1$. This is reflected in Figure~\ref{fig:shock_front} as the precursor wave is barely visible in the $\lambda = 0.2 r_{\rm nl}$ case.

\paragraph{Application to astrophysical FRBs.---}

We consider a fast wave with a luminosity of $L = L_{42}10^{42}\, {\rm erg \, s^{-1}}$, to be consistent with typical magnetar X-ray bursts. Its frequency is determined by the underlying starquake, on the order of $\omega=\omega_4 10^4\,{\rm rad\,s}^{-1}$~\cite{1989ApJ...343..839B}. Suppose the magnetar has a surface magnetic field strength of $B_* = B_{15}10^{15}\, {\rm G}$ and a spin period of $P = P_0 1\, {\rm s}$. We estimate the plasma density at the stellar surface to be $n_* \approx \mathcal{M} n_{\rm GJ}\approx 6.9\times10^{19}\,B_{15}\mathcal{M}_6 P_0^{-1} {\rm cm}^{-3}$ where $\mathcal{M} = \mathcal{M}_6 \, 10^6$ is the pair multiplicity and $n_{\rm GJ} = B_*/ecP = 6.9\times10^{13}\,B_{15}P_0^{-1}\,{\rm cm}^{-3}$ is the Goldreich Julian density~\cite{1969ApJ...157..869G} at the stellar surface. These parameters and Equation~\eqref{eq:r_nl} imply a nonlinear radius of $r_{\rm nl}\approx2.9\times10^8 \,B_{15}^{1/2} L_{42}^{-1/4} \,{\rm cm}$ and the ratio of the fast wave wavelength to the nonlinear radius is:
\begin{equation}
    \lambda/r_{\rm nl} \approx 6.5\times10^{-2} \, B_{15}^{-1/2} L_{42}^{1/4}  \omega_4^{-1}.
\end{equation}
The background magnetization at the non-linear radius is then:
\begin{equation}
    \sigma_{\rm bg,nl} \equiv \frac{B_\mathrm{bg,nl}^{2}}{4\pi n_\mathrm{bg, nl}m_{e}c^{2}} = 5.7\times10^7 \,B_{15}^{-1/2} L_{42}^{3/4}M_6^{-1}P_0.
\end{equation}
Using Equation~\eqref{eq:gamma_max}, the maximum value that the upstream Lorentz factor will attain is
\begin{equation}
    \gamma_{\rm max}
    \sim 7.7\times10^5 \, B_{15}^{-1} L_{42} M_6^{-1}P_0 \omega_4^{-1}.
\end{equation}
This happens just slightly past the nonlinear radius where the shock first forms. The upstream magnetization is reduced to, according to Equation~\eqref{eq:sigma_u},
\begin{equation}
    \sigma_{\rm u}(r_{\rm nl})\approx \pi r_{\rm nl}/(1.3\lambda)\approx 37 \,B_{15}^{1/2} L_{42}^{-1/4}\omega_4.
\end{equation}

For definiteness, we estimate the properties of the expected radio emission at the magnetic equator, since the obliquity of the shock makes the prediction about precursor wave emission much more difficult elsewhere. According to \cite{Plotnikov2019}, the low frequency cutoff of the precursor wave in the downstream frame is $\omega_{\rm cutoff} = \gamma_{\rm sh|d}\omega_p \approx \sqrt{\sigma_{u}}\omega_p$ and the peak frequency is $\omega_{\rm peak} \approx 3\omega_{\rm cutoff}$, where $\omega_p = \sqrt{4\pi ne^2/m_e \gamma}$ is the proper plasma frequency of the upstream measured in the downstream frame. When the shock first forms, the upstream Lorentz factor reaches its maximum, $\gamma \approx \gamma_{\rm max}$, the magnetic field, density, and magnetization reduce to half their background value. Combining these, the cutoff frequency immediately after the shock forms is $\omega_{\rm cutoff} \approx 4.7\times10^8 \, B_{15}^{1/2}L_{42}^{-1/4}M_6P_0^{-1}\omega_4 \,{\rm rad\,s}^{-1}$. Therefore, the precursor wave spectrum will peak near:
\begin{equation}
    \nu_{\rm peak}(r_{\rm nl})=\frac{\omega_{\rm peak}(r_{\rm nl})}{2\pi} \approx 0.22 \, B_{15}^{1/2}L_{42}^{-1/4}M_6P_0^{-1}\omega_4 \,{\rm GHz}.
\end{equation}
The precursor wave efficiency, defined as the fraction of the total incoming energy entering the shock that is channeled into the precursor waves, is $f\approx2\times 10^{-3}/\sigma_{u}$~\cite{Plotnikov2019}. In our case, the precursor wave luminosity when the shock first forms just beyond $r_{\rm nl}$ is:
\begin{equation}
    L_R(r_{\rm nl})\sim fL\sim 5.4\times10^{37}B_{15}^{-1/2}L_{42}^{3/4}\omega_4^{-1}\,{\rm erg\,s}^{-1}.
\end{equation}

With appropriate parameters, the precursor wave frequency and luminosity can be consistent with observed FRBs, especially FRB 200428 from the galactic magnetar SGR 1935+2154. FRB 200428 was detected by CHIME, with an energy emitted in the 400-800-MHz band of $3\times 10^{34}\,\rm{erg}$ and a peak luminosity of $7\times 10^{36}\,\rm{erg\,s^{-1}}$~\cite{2020Natur.587...54C}. It was also detected by STARE2 in the $1.281-1.468$~GHz band, with an isotropic equivalent energy release of $2.2\times 10^{35}\,\rm{erg}$ and a full-width at half-maximum (FWHM) temporal width of $\sim0.6\,\rm{ms}$, suggesting a luminosity $\sim4\times 10^{38}\,\rm{erg\,s^{-1}}$~\cite{2020Natur.587...59B}. Since SGR 1935+2154 has a surface magnetic field of $B_{15}\sim 0.2$ and a period $P_0\approx 3.2$, we can see that $\nu_{\rm peak}(r_{\rm nl})$ would be around 1.4~GHz if we take $M_6\sim45$, $\omega_4\sim1$ and $L_{42}\sim1$; these parameters would give $L_R(r_{\rm nl})\sim10^{38}\,\rm{erg\,s^{-1}}$, consistent with observations obtained by CHIME and STARE2. FRB 200428 was also accompanied by an X-ray burst, with a duration of $\sim0.6$~s, superimposed by spikes coincident with the radio bursts. The energy released in the X-ray burst is $\sim 10^{40}\,\rm{erg\,s^{-1}}$, consistent with our choice of a fast wave luminosity of $L_{42}\sim1$~\cite{2020ApJ...898L..29M, 2021NatAs...5..372R,2021NatAs...5..378L}.

As the shock propagates to larger radii, the upstream Lorentz factor evolves as $\gamma\sim \gamma_{\rm max}(r/r_{\rm nl})^{-4}$ according to Equation~\eqref{eq:full_monster}, and the upstream magnetic field is kept at $B_{\rm up}=B_{\rm bg}/2\propto r^{-3}$, while the density remains as $n=n_{\rm bg}/2\propto r^{-3}$, therefore, the upstream magnetization evolves as $\sigma_{u}(r)\propto r$. The upstream proper plasma frequency evolves as $\omega_p\propto \sqrt{n/\gamma}\propto \sqrt{r}$. This results in the peak frequency of the precursor wave increasing with radius: $\omega_{\rm peak}\approx3\sqrt{\sigma_u}\omega_p\propto r$. However, the luminosity of the precursor waves quickly drops with radius: the total incoming power entering the shock front is $L_u\sim r^2cB_{\rm up}^2\propto r^{-4}$, therefore, $L_R\sim f L_u\propto r^{-5}$. The luminosity of the precursor waves reduces to less than $0.5\%$ of its peak value $L_R(r_{\rm nl})$ when the shock has propagated to a radius of $r=3r_{\rm nl}$. All of the radio emission is effectively produced between $r_\mathrm{nl} < r < 3r_\mathrm{nl}$.

We also estimate the duration of the observed precursor wave emission. We take the duration of the emission as $\Delta t\sim 2r_{\rm nl}/c$, and the shock Lorentz factor is approximately $\gamma_{\rm sh}\sim\sqrt{\sigma_u(r_{\rm nl})}$ (a lower bound), so we get the observed duration $\Delta t_{\rm obs}\sim \Delta t/\gamma_{\rm sh}^2\sim 0.5\omega_4^{-1}\,{\rm ms}$ (an upper bound). This is shorter than typical FRBs. However, here we only considered the first shock launched in the first wavelength of the fast wave. The subsequent wavelengths of the fast wave can also launch shocks when they become nonlinear; the downstream of the previous shock will become the upstream of the next shock. Although the precursor wave emission can be suppressed when the upstream plasma is relativistically hot~\cite{2020MNRAS.499.2884B}, if the cooling of the plasma is efficient (e.g., through incoherent synchrotron radiation), the subsequent shocks could have a sufficiently cold upstream that allows precursor wave emission. The total energetics and duration of the observed radio emission will then be determined by the full fast wave train, and the individual shocks may produce substructures in the observed bursts, on a time scale of $\delta t\sim \lambda/c\sim 0.6\omega_4^{-1}\,{\rm ms}$. This time scale may correspond to the observed substructure in some bursts~\cite{2021ApJ...919L...6M,2022Natur.607..256C}. Since our simulations do not include any cooling processes, we cannot properly model the behavior of subsequent shocks. We leave this to future studies.

The monster shock scenario may not be able to account for the very bright cosmological FRBs. For example, the bright FRB 20220610A has a luminosity $L_R\sim 10^{45}\,{\rm erg\,s}^{-1}$~\cite{2023Sci...382..294R}. This would require a fast wave with luminosity $L\sim 10^{48}\,{\rm erg\,s}^{-1}$, which would become nonlinear at a radius $r_{\rm nl}\sim9.3\times 10^6\,{\rm cm}$. Such a level of energy release in the highly magnetic compact region close to the magnetar would lead to copious pair production and the plasma would quickly become an optically thick fireball, similar to the giant flares~\cite{1995MNRAS.275..255T}. It is questionable whether any radio waves can be produced or escape from such an environment. Alternative explanations for these bright FRBs are still needed.

The monster shock can also produce incoherent radiation in other wavelengths. Immediately downstream of the shock, the energetic particles entering from the upstream start to gyrate in the compressed magnetic field and produce synchrotron/curvature radiation.
If we adopt the naive synchrotron formula, we get the maximum photon energy $E_{\rm syn} =3h\gamma_{\rm max}^2e B_{\rm bg}/(4\pi m_ec)\sim0.4\, B_{15}^{-5/2} L_{42}^{11/4} M_6^{-2}P_0^2 \omega_4^{-2} \, {\rm TeV}$.
Note that this energy is comparable to $\gamma_{\rm max}m_e c^2$, indicating that the synchrotron emission likely proceeds in the quantum regime, where the particle will quickly give most of its energy to one single photon. The gamma ray photons at this energy can already produce pairs through magnetic pair production, increasing the downstream plasma density. A more careful radiative transfer calculation is required to compute the detailed pair yield and the resulting X-ray spectrum, and will be deferred to a future work.

\clearpage

\section{Supplemental Material}
\setcounter{page}{1}
\subsection{Simulation setup}

We use the GPU-accelerated PIC code \emph{Aperture} designed for studying relativistic plasmas around compact objects~\footnote{https://github.com/fizban007/Aperture4}. The code solves the Maxwell-Vlasov system using the standard Particle-in-Cell algorithm on a spherical grid with logarithmic spacing in the $r$ direction. The electromagnetic field is interpolated to the particle position using a first order spline, then used to update the particle momenta using the Vay pusher~\cite{2008PhPl...15e6701V}. The motion of the particle is used to deposit a current density onto the grid with a first order shape function using a modified Esirkepov method that conserves charge~\cite{2001CoPhC.135..144E,2025arXiv250304558C}. Finally, the current is used to update the electromagnetic fields using a fully explicit second order scheme based on the integral version of the Maxwell equations, similar to~\cite{2015MNRAS.448..606C}. The code has been successfully applied to perform global simulations of the neutron star magnetosphere in multiple works in the past decade~\cite{2014ApJ...795L..22C, chen-thesis, 2017ApJ...844..133C, 2020ApJ...889...69C}.

The simulations described in this work begin with a non-rotating, unperturbed dipolar magnetosphere filled with a cold pair plasma whose number density is $n_{\rm bg}=N_0/r^3$, where $N_0$ is a constant. This is achieved by setting a radius-dependent weight on each particle, but maintaining a constant number of particles per cell through out the simulation domain. We typically have 12 particles per cell in the initial plasma. Beyond this initial background plasma, no additional pairs are added throughout the simulation. The inner boundary of our simulation domain is located at $r_0 = 0.4r_{\rm nl}$, where $r_{\rm nl}$ is the radius at which the fast wave becomes nonlinear, namely $\delta B/B_{\rm bg} \approx 1/2$. The nonlinear radius $r_{\rm nl}$ is a natural length unit of the problem. Since the fast wave propagation is linear up to $r_{\rm nl}$, we choose the inner boundary $r_0 = 0.4r_{\rm nl}$ to focus our attention on the dynamics near the nonlinear radius. At $r_0$, the boundary condition for particles is such that any particles crossing the inner boundary are absorbed. Such a boundary condition has no effect on the behavior of the fast wave at the nonlinear radius. Fast waves are launched into the magnetosphere by imposing an oscillating toroidal electric field $E_{\phi}$ at the inner boundary of the
simulation domain, $r = r_0$:
\begin{equation}
    \label{eq:launching}
    E_{\phi}(r_0, \theta, t) = E_0\,\sin\theta \sin(\omega t),
\end{equation}
where $\theta$ is the spherical polar angle, and $\omega$ is the angular frequency of the fast wave. Such waves can be produced either directly from the starquakes~\cite{Beloborodov2023,2025arXiv250812567Q,2025arXiv250818033B} or through the spontaneous conversion of Alfv\'{e}n waves in the inner magnetosphere~\cite{2021ApJ...908..176Y,2024ApJ...972..139M,2025ApJ...980..222B,2025ApJ...987...42C}. At a radius $r_0=0.4r_{\rm nl}\gg r_*$ (see Equation \ref{eq:r_nl} for typical values), the waves are already propagating as asymptotic spherical waves, agnostic of the launching mechanisms. The $\sin\theta$ profile is chosen as it represents the lowest (dipole) order in the radiation multipole expansion~\cite{jackson1999classical}, and it also conveniently avoids any pathological behavior near the axis. We simulated fast waves with wavelength $\lambda = 2\pi c/\omega$ ranging from $0.2r_{\rm nl}$ to $r_{\rm nl}$. The outer boundary is chosen near $r_1 \approx 4.7r_{\rm nl}$. We apply an open boundary condition at this radius, but since we typically terminate our simulations before the fast wave reaches the outer boundary, the exact boundary condition implementation does not play a significant role.

To properly capture the shock physics, the simulations need to resolve both the plasma skin depth $\lambda_p=c/\omega_p=\sqrt{m_ec^2/(4\pi n_{\rm bg} e^2)}$  and the gyration time scale $\omega_B^{-1}=m_ec/(eB_{\rm bg})$ at the nonlinear radius. The ratio of these two scales is the magnetization, which is a key parameter of this problem:
\begin{equation}
\sigma_{\rm bg,nl} = \frac{B_{\rm bg}^2}{4\pi n_{\rm bg}m_e c^2} = \left(\frac{\omega_B}{\omega_p}\right)^2.
\end{equation}
In our simulations, we typically choose $\lambda_p \sim 2\times 10^{-3}\,r_{\rm nl}$ by adjusting the normalization constant $N_0$, and we have performed simulations with $\sigma_{\rm bg,nl}$ ranging from $160$ to $2000$. Our grid resolution ranges from $6144\times 8192$ to $12288\times 16384$, uniformly spaced in $\ln r$ and $\theta$. These resolutions ensure that we have at least 4 cells
per skin depth at the non-linear radius. 

\subsection{Condition for shock development}

Monster shocks behave according to MHD in a sufficiently dense plasma. In the vacuum limit however, the wave should not deform, and instead propagate as a vacuum electromagnetic wave. \cite{2022arXiv221013506C}~studied the regime in which the steepening of fast waves transitions from being well described by MHD to a regime where it is not. It was found that as long as the magnetosphere satisfies (in a dipole background field)
\begin{equation}
    \label{eq:shock-condition}
    \eta = \omega_p^2 r_{\rm nl}/c\omega_B \gtrsim 10,
\end{equation}
the plasma behaves like in MHD and the wave steepens into a shock. The reason for this criterion is that $E_\phi$ is reduced by a toroidal current $j_\phi$, which is produced by the accelerated upstream plasma. As the plasma is accelerated in the $-\hat{r}$ direction, the positive and negative charges are accelerated to $+\hat{\phi}$ and $-\hat{\phi}$ directions respectively, providing a toroidal current. Since the toroidal velocities asymptote to $\sim\pm 0.1c$, a high enough upstream plasma density is required to provide enough $j_\phi$ to screen the wave electric field, leading to Equation~\ref{eq:shock-condition}.

\begin{figure}[h]
    \centering
    \includegraphics[width=0.99\linewidth]{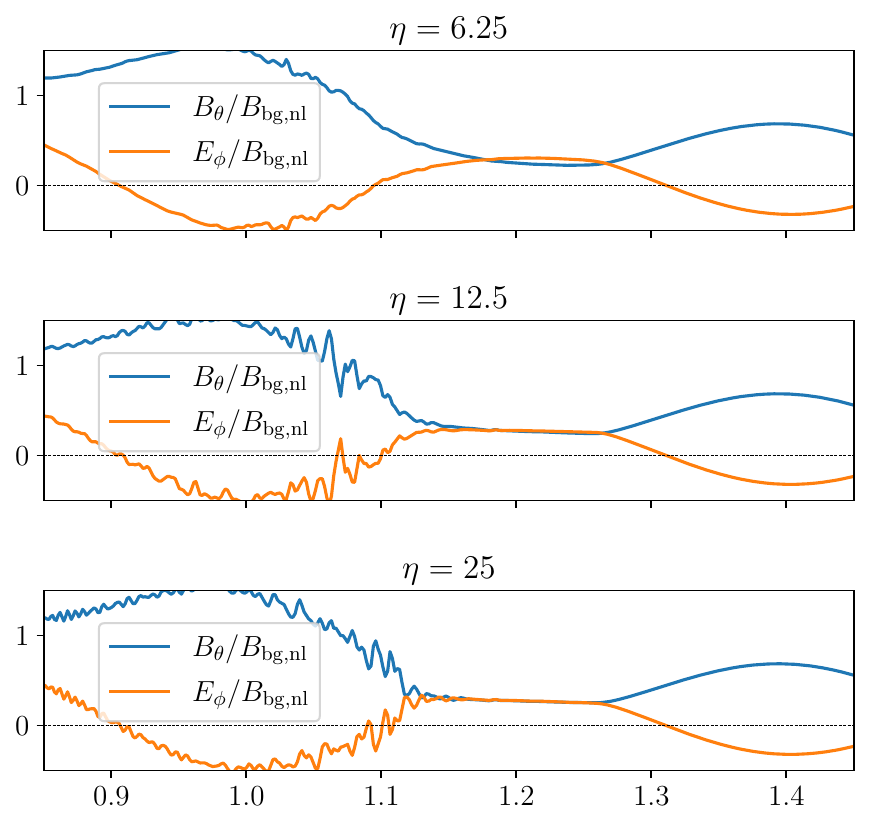}
    \caption{
    Snapshots of the electric and magnetic fields from three simulations with different $\eta$ parameters, demonstrating different regimes of wave deformation. Each snapshot is taken at the timestep $t = 1.1r_{\rm nl}/c$. All three simulations have the same background magnetic field and wavelength. The only variable that changes is the plasma density. The top panel shows $\eta = 6$ to demonstrate the regime in which we do not expect $E>B$ to be successfully prevented. The center panel shows $\eta = 12.5$ which is close to the threshold necessary to prevent $E>B$. The bottom panel shows $\eta = 25$ in which $E>B$ is completely prevented with the exception of the precursor wave.}
    \label{fig:eta}
\end{figure}

\begin{figure}[ht!]
    \centering
    \includegraphics[width=0.99\linewidth]{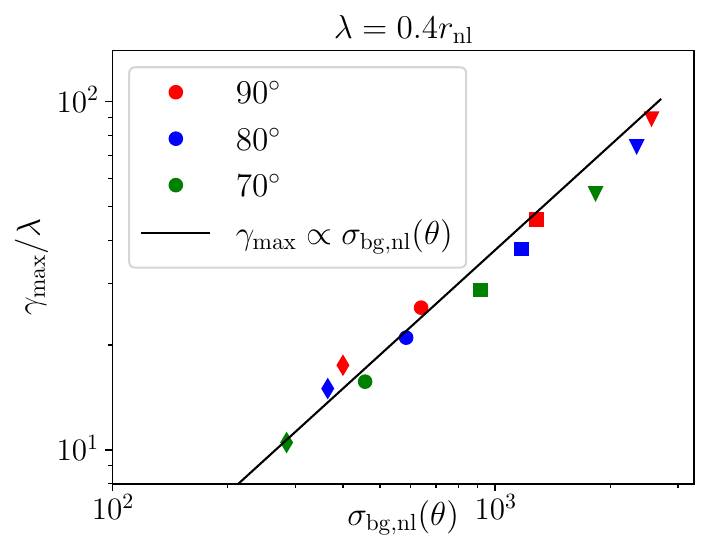}
    \caption{The maximum value that the upstream bulk Lorentz factor attains at several different poloidal angles. The wavelength is constant at $\lambda = 0.4r_{\rm nl}$ for all simulations, but we still divide $\gamma_{\rm max}$ by $\lambda$. Consequently, each red marker in this figure corresponds to a point in Figure~\ref{fig:Lorentz}. Distinct simulations are indicated by distinct markers styles, while colors are consistent with distinct angles. The only simulation that does not satisfy MHD is plotted as triangles.}
    \label{fig:Lorentz_2D}
\end{figure}

We test this prediction using simulations with several different values of $\eta$. Figure~\ref{fig:eta} shows the electric and magnetic field shortly after the wave becomes non-linear for simulations with different values of $\eta$. In the case of $\eta = 25$, the fast wave steepens and a shock forms. With the exception of the early stages of the precursor wave, $E>B$ is prevented as~\cite{2022arXiv221013506C} predicted. For the boundary case, $\eta = 12.5$, $E>B$ is almost completely prevented in the upstream region, but the shock barely forms. In the case of $\eta<10$, $E>B$ occurs, violating the MHD assumption. The wave has only deformed slightly and a shock has not formed. In the region of $E>B$, the plasma is accelerated to very high Lorentz factors, similar to the upstream of a monster shock. The simulations with the highest value of $\sigma_{\rm bg, nl}$ in Figure~\ref{fig:Lorentz} are the only two simulations where we see $E>B$ occurring. Despite not forming a monster shock, the plasma Lorentz factor resulting from the $E > B$ region still follows the same $\gamma_{\rm max}\propto \sigma$ scaling. For the magnetar and fast wave parameters as discussed in the main text, we have $\eta = 1.2\times10^5\,B_{15}^{1/2}L_{42}^{-1/4}M_6P_0^{-1}$ which is well within the MHD regime.

We discussed previously the formation of regions with $E>B$ in Figure \ref{fig:shock_front} at higher latitudes for later wavelengths. This is because the plasma bulk velocity at these locations is directed toward the equator, increasing the density near the equator while reducing the density at higher latitudes. As a result, the $\eta$ parameter drops below 10 in the high latitude regions, and $E>B$ can no longer be prevented. Our simulations did not include any pair production, though. In a realistic magnetar magnetosphere, such macroscopic regions of $E>B$ could lead to particle acceleration and copious pair production, so that condition (\ref{eq:shock-condition}) could be maintained and $E$ would remain less than $B$.

\subsection{Scaling of the upstream Lorentz factor away from the equator}

We also study the scaling of the upstream Lorentz factor outside of the equatorial plane. To demonstrate this, we measure the maximum value that the upstream Lorentz factor attains at different poloidal angles for several different simulations. The simulations we consider span a range of values of $\eta$, primarily focusing on the regime where MHD is satisfied. Note that for different polar angles, the radius where the fast wave first becomes nonlinear is different. We write the angular dependent nonlinear radius as $r_{\rm nl}(\theta)$. To be consistent with the rest of the text, we continue to normalize everything to the nonlinear radius in the equatorial plane $r_{\rm nl, eq}$ and any value not restricted to the equatorial plane will be explicitly written as a function of $\theta$. Before the wave deforms, the wave magnetic field at the wave trough is given by $B_{\rm w}(\theta) = -(B_{\rm nl, eq}/2)(r_{\rm nl,e   q}/r)\sin\theta\,\hat{\theta}$ and the electric field is given by $E_{\rm w}(\theta) = (B_{\rm nl,eq}/2)(r_{\rm nl,eq}/r)\sin\theta\,\hat{\phi}$. The total magnetic field is then
\begin{equation}
    B(\theta) = \frac{B_{\rm nl, eq} r_{\rm nl, eq}^3}{r^3}\left(2\cos\theta\,\hat{r} + \left(1-\frac{r^2}{2r_{\rm nl, eq}^2}\sin\theta\right)\hat{\theta}\right).
\end{equation}
Taking $E(\theta)^2 - B(\theta)^2 = 0$ and solving for the radius as a function of $\theta$ gives the non-linear surface 
\begin{equation}
    r_{\rm nl}(\theta)=r_{\rm nl,eq}\sqrt{\frac{4-3\sin^2\theta}{\sin^2\theta}}.
\end{equation}
Figure~\ref{fig:Lorentz_2D} shows the maximum value of the upstream Lorentz factor $\gamma_{\rm max}(\theta)$ as a function of the magnetization at the non-linear radius for the corresponding poloidal angle $\sigma_{\rm bg, nl}(\theta)$. Individual simulations are indicated by distinct markers. For a fixed poloidal angle, $\gamma_{\rm max}(\theta)$ depends linearly on $\sigma_{\rm bg, nl}(\theta)$.

\subsection{Limitations and Outlook}
Monster shocks are a natural consequence of a dynamic magnetosphere and can produce coherent radio emission. Consequently they have recently been studied in detail in 1D~\cite{2025PhRvL.134c5201V}. However, this is fundamentally a multi-dimensional problem, resulting in a shock with non-trivial global structure that will affect the overall properties of both the precursor emission, and the shock propagation. This paper examined the formation of the monster shock in a realistic background magnetic field configuration, and showed the latitude dependence of the shock structure. In particular, using a realistic dipolar background magnetic field as well as adopting the correct spherical geometry allowed us to directly measure e.g.\ the evolution of upstream Lorentz factor over time, which is crucial for predicting the properties of precursor waves.

Since the main improvement of our model over previous 1D studies is the inclusion of dipole background field and spherical wave propagation, we expect similar conclusions to hold when going from 2D axisymmetry to full 3D. This is because as long as the fast wave is generated close to the star, regardless of the mechanism, after spherical propagation to the nonlinear radius $r_\mathrm{nl}$ the wave vector becomes approximately radial. Additionally, when the size of the emitting region is smaller than the wavelength, classical radiation multipole expansion applies and the wave profile at large distances is dominated by the dipole component considered in this paper. 3D simulations reported by~\cite{2022ApJ...933..174Y} indeed showed that the waves launched from a small region on the stellar surface behave in a similar way as their axisymmetric counterparts except near the boundaries of the wave profile. Future 3D global simulations of the monster shock may be able to study the polarization of the precursor emission, especially its dependence on the viewing angle.

Another important physical process neglected in these simulations is strong radiative cooling and the resulting pair production. As mentioned above, the photons produced from synchrotron cooling of the shock heated plasma can produce pairs. Because monster shocks from kHz waves are highly relativistic, the pairs produced in the downstream will not be able to catch the shock. However, these pairs can increase the plasma density downstream of the shock, and affect the evolution of the later part of the fast wave train. Moreover, strong cooling will affect the structure of the solitons and consequently affect the precursor emission. Recent simulations have suggested that this may result in more narrow band emission about the peak frequency, strengthening the association with FRBs~\cite{Zhang_2025}. However a detailed study of the precursor emission from monster shocks in the presence of strong cooling is necessary but outside of the scope of this paper. In addition, strong cooling will become significant for later wavelengths in the fast wave wave train. With strong cooling, subsequent wavelengths will encounter a cold plasma and shock, similar to the first wavelength. However the density will be significantly increased from pair production and the plasma that flows in from higher latitudes after the earlier shocks pass through. This will reduce the background magnetization and increase the plasma frequency, significantly effecting the multiwavelength emission of each successive shock.

\end{document}